\documentclass{PoS}
\newcommand{\gsim}{\raisebox{-0.07cm   }
{$\, \stackrel{>}{{\scriptstyle\sim}}\, $}}

\usepackage{amsmath}
\usepackage{float}
\usepackage{cite}

\title{{\footnotesize DESY 13-133,~~~DO-TH 13/16,~~~MITP/13-038,~~~SFB/CPP-13-51,
~~~LPN 13-053}\\
Recent Results on the 3-Loop Heavy Flavor Wilson 
               Coefficients in Deep-Inelastic Scattering\thanks{
This work has been supported in part by DFG Sonderforschungsbereich
Transregio 9, Computergest\"utzte Theoretische Teilchenphysik, by Studienstiftung des Deutschen
Volkes, by the Austrian Science Fund (FWF) grants P20347-N18, P22748-N18 and SFB F50
(F5009-N15), by the EU Network {\sf LHCPHENOnet} PITN-GA-2010-264564, by the Reserach Center 
`Elementary Forces and Mathematical Foundations (EMG) of J. Gutenberg University Mainz and DFG, 
and by FP7 ERC Starting Grant  257638 PAGAP.}
}

\ShortTitle{3-Loop Heavy Flavor Wilson Coefficients}

\author{\speaker{J. Bl\"umlein}, A. De Freitas, C. Raab, F. Wi\ss{}brock 
\\
        Deutsches Elektronen-Synchrotron DESY, Platanenallee 6, D-15738 Zeuthen, Germany\\
        E-mail: \email{Johannes.Bluemlein@desy.de}}

\author{J. Ablinger, A. Hasselhuhn, M. Round, C. Schneider\\
Research Institute for Symbolic Computation (RISC),\\
                          Johannes Kepler University, Altenbergerstra\ss{}e 69,
                          A--4040, Linz, Austria}

\author{A. von Manteuffel\\
        PRISMA Cluster of Excellence and Institute of Physics, J. Gutenberg University, 
D-55099 Mainz. Germany.}

\abstract{We report on recent progress in the calculation of the 3-loop massive Wilson 
coefficients in deep-inelastic scattering at general values of $N$ for neutral and charged 
current reactions in the asymptotic region $Q^2 \gg m^2$.}

\FullConference{XXI International Workshop on Deep-Inelastic Scattering and Related 
                Subjects -DIS2013,\\ 22-26 April 2013\\ Marseilles,France}

\begin{document}
%-------------------------------------------------------------------------------------------------
%-------------------------------------------------------------------------------------------------
\section{Introduction}
%-------------------------------------------------------------------------------------------------

\noindent
The precision determinations of $\alpha_s(M_Z^2)$, the mass of the charm quark $m_c$ and the 
parton distribution functions from the world data on deep-inelastic scattering (DIS) require 
the heavy flavor corrections to 3-loop order \cite{Alekhin:2012vu}. Here the structure function
$F_2(x,Q^2)$ provides the highest precision. As has been shown in \cite{Buza:1995ie} at scales
$Q^2/m_c^2 \gsim 10$ the asymptotic representation of the heavy flavor Wilson coefficients
provides a representation on the per cent level.~\footnote{The corresponding scales are much 
higher in case of the structure function $F_L(x,Q^2)$ \cite{Buza:1995ie}, for which the 
3-loop heavy flavor corrections for general values of $N$ have been calculated in 
\cite{Blumlein:2006mh}.} They are given in terms of convolutions of massive operator matrix
elements (OMEs) and the massless Wilson coefficients \cite{Vermaseren:2005qc}. A series 
of 3-loop Mellin-moments for $F_2(x,Q^2)$ and transversity and the OMEs describing the
transition matrix elements in the variable flavor number scheme (VFNS) 
\cite{Buza:1996wv,Bierenbaum:2009zt}
have been calculated in 2009 in Refs.~\cite{Bierenbaum:2009mv,Blumlein:2009rg} projecting the 
respective tensor quantities onto massive tadpoles which could be computed using {\tt MATAD} 
\cite{Steinhauser:2000ry}. 
 
A program to compute the massive 3-loop Wilson coefficients at general values of $N$ and 
their analytic continuation to $N \in \mathbb{C}$ started thereafter. In the unpolarized case,
eight Wilson coefficients/OMEs contribute. All logarithmic contributions  
\cite{Bierenbaum:2010jp} are available since they rely on the the 2-loop results 
\cite{Buza:1995ie,Bierenbaum:2007qe} up to $O(\alpha_s^2 \varepsilon)$ \cite{Bierenbaum:2008yu}.
Two of the eight Wilson coefficients resp. OMEs, $L_{qg,Q}^{(3)}$ and $L_{qq,Q}^{(3),\rm PS}$, 
were calculated in \cite{Ablinger:2010ty}. We studied the contributions to specific color factors,
such as $O(N_F T_F^2 C_{A,F})$, which are completely known now 
\cite{Ablinger:2010ty,Blumlein:2012vq}. 
Further investigations are devoted to diagrams with two fermion lines with finite equal 
\cite{Ablinger:2012ej} or unequal mass \cite{Ablinger:2011pb,BW13}. Genuine 3-loop topologies of 
the ladder- and V-graph type have been studied in \cite{Ablinger:2012sm,Ablinger:2012qm}. These 
calculations were accompanied by mathematical and computer-algebraic developments. In course of 
this systematic use is made of higher hypergeometric functions, Mellin-Barnes techniques, 
and modern summation theory \cite{SIGMA}. The latter are encoded in the packages {\tt Sigma, 
EvaluateMultiSums} and {\tt SumProduction} \cite{CODE1}. Extensions of the harmonic sums 
\cite{HSUM} and polylogarithms \cite{Remiddi:1999ew} to generalized harmonic sums 
\cite{Moch:2001zr,Ablinger:2013cf} and the associated iterated integrals, the cyclotomic and 
generalized cyclotomic sums and integrals \cite{Ablinger:2011te} were developed. Most recently 
iterated integrals over root-valued letters were systematized. These functions and their relations 
were encoded in the package {\tt HarmonicSums}, \cite{Ablinger:2013cf,Ablinger:2013hcp}, see also 
\cite{Ablinger:2013jta}. All these developments were necessary to perform the present 
calculations. They are, however, of much wider use.
   
In this note we report on progress being obtained during the last year.
%-------------------------------------------------------------------------------------------------
\section{3-Loop OMEs with Two Fermion Lines of Equal Mass}
%-------------------------------------------------------------------------------------------------

\begin{figure}[t]
\begin{center}
\includegraphics[scale=0.5]{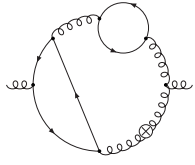}
\end{center}
\caption[]{An example for a graph with two massive fermion lines}\label{FIG3}
\end{figure}

\noindent
A subset of graphs contributing to the 3-loop massive Wilson coefficients contains two fermion
lines with equal mass, characterized by the color factor $T_F^2 C_{F,A}$. These graphs may contain
new types of sums, which, to a wider extent also emerge in the V-topologies, see 
Section~\ref{sec:BV}. 
These are weighted inverse binomial sums. An example is given by the diagram in Figure~\ref{FIG3}.
%-------------------------------------------------------------------------------------------------
%-------------------------------------------------------------------------------------------------
The diagram is given by 
%-------------------------------------------------------------------------------------------------
\begin{eqnarray}
I(N) &=& \frac{1 + (-1)^N}{2} \Biggl\{
\frac{1}{45 \varepsilon^2 (N+1)} 
- \frac{1}{\varepsilon} \left[\frac{S_1(N)}{90(N+1)}
+\frac{47 N^3 + 20 N^2 -67 N +40}{1800(N-1)N(N+1)^2}\right]
\nonumber\\
&&
+\frac{105 N^3 - 175 N^2 + 56 N + 96}{13440(N+1)^2(2N-3)(2N-1)4^N}\binom{2N}{N}\left[
\sum_{j=1}^N \frac{4^j S_1(j)}{\binom{2j}{j} j^2} - \sum_{j=1}^N \frac{4^j}{\binom{2j}{j} 
j^3} - 7 \zeta_3\right]
\nonumber\\ &&
+ \frac{5264 N^3 -2409 N^2 -12770 N +3528}{100800(N+1)^2(2N-3)(2N-1)} S_1(N)
+ \frac{S_1^2(N)+S_2(N) +3 \zeta_2}{360(N+1)}
\nonumber\\ &&
+ \frac{S_3(N) - S_{2,1}(N) + 7 \zeta_3}{420(N+1)}
+ \frac{Q_0(N)}{2268000(N-1)^2N^2(N+1)^3(2N-3)(2N-1)}
\Biggr\}~.
\nonumber
\end{eqnarray}
%-------------------------------------------------------------------------------------------------
Here and in the following $Q_i$ denote polynomials in $N$.
The terms $\propto 1/(2N-3), 1/(2N-1)$ deserve special attention. It can be shown that both are 
removable poles in $I(N)$. It is generally expected that in QCD the rightmost singularity is 
located at $N=1$. All basic topologies of this type contributing to the OME $A_{gg}^{(3)}$ have 
been calculated.
%-------------------------------------------------------------------------------------------------
\section{3-Loop OMEs with Two Fermion Lines of Different Mass}
%-------------------------------------------------------------------------------------------------

\noindent
From the level of the 3-loop correction onwards, also graphs with two fermion lines of different 
mass contribute. They require an extension of the renormalization programme of 
Ref.~\cite{Bierenbaum:2009mv}. It turns out that the equal mass case is better included alongside 
with the case of two different masses $m_c$ and $m_b$. The very close values of the charm and bottom 
quark masses do not allow to treat charm massless at the scale $\mu^2 = m_b^2$ and one has 
to deal with a two-mass scenario. Yet $\xi =  m^2_c/m_b^2 \sim 1/10$ allows an expansion in $\xi$.
For the fixed moments $N = 2,4,6$ the calculation of all OMEs has been performed in 
\cite{Ablinger:2011pb,BW13} after 
mapping them to tadpoles and using the code {\tt qexp} \cite{QEXP}. First results were 
derived for general values of $N$.
It is needless to say that also the matching conditions in the variable flavor scheme require 
these new and no other expressions to stay in accordance with the renormalization group 
equations inside the correct framework of perturbative QCD. Moreover, the matching scales 
may vary considerably for different observables \cite{Blumlein:1998sh}. 
%-------------------------------------------------------------------------------------------------
\restylefloat{figure}
\begin{figure}[h]
\begin{center}
\includegraphics[scale=0.8]{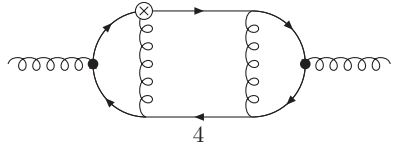} 
\caption[]{Ladder graph with operator insertion.}\label{FIG1}
\end{center}
\end{figure}
%-------------------------------------------------------------------------------------------------
\section{Ladder Graphs}
%-------------------------------------------------------------------------------------------------

\noindent
First results have been obtained in the calculation of ladder graphs in the massive case, which 
belong to the genuine 3-loop topologies \cite{Ablinger:2012qm}. Here the class of functions 
appearing in intermediate and final results extends to generalized harmonic sums, 
cf.~\cite{Ablinger:2013cf}. Let us consider the diagram in Figure~\ref{FIG1}. 
The corresponding scalar graph yields
%-------------------------------------------------------------------------------------------------
\begin{center}
\includegraphics[angle=0,width=\textwidth]{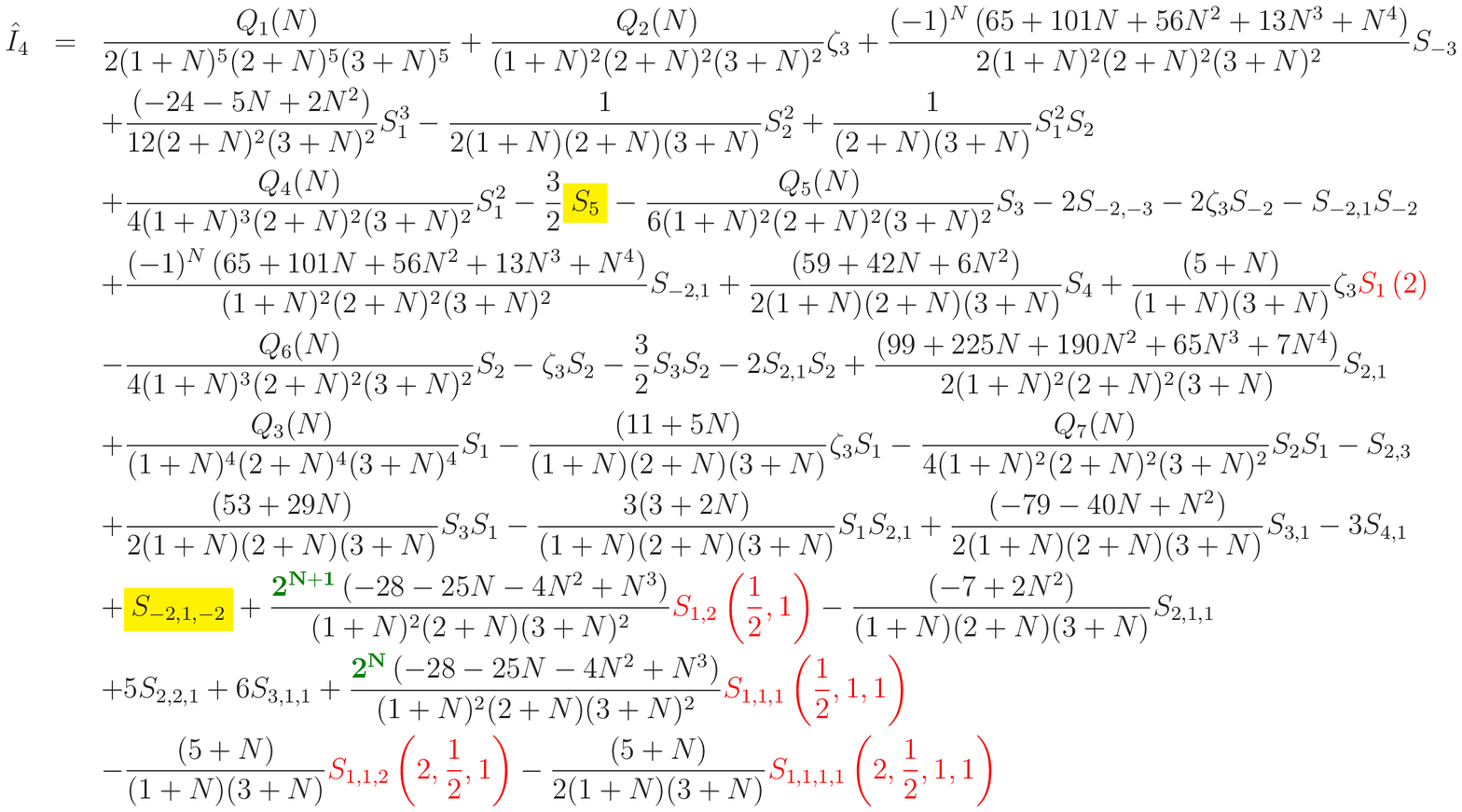}
\end{center}
%-------------------------------------------------------------------------------------------------
It can be calculated with an extension of the method of hyperlogaritms \cite{Brown:2008um}
to the case of massive graphs with operator insertion \cite{Ablinger:2012qm} and is of weight {\sf 
w = 5}. One notices the 
emergence of terms growing individually like $\propto 2^N$, which would potentially imply an 
instability at large $N$. However, the asymptotic expansion of the function $\hat{I}_4(N)$
shows that the corresponding terms cancel. In case of this and more involved  topologies 
both in the sum-representation and likewise also in that by iterated integrals the individual 
entities of the representation, despite spanning the algebraic basis, partly act together forming 
the physical structures. Individually they may not reflect the properties of the complete diagram. 

%-------------------------------------------------------------------------------------------------
\section{Massive Benz and V-Topologies}
\label{sec:BV}
%-------------------------------------------------------------------------------------------------

\noindent
The method of hyperlogarithms is also suited to compute non-divergent diagrams of other 
massive topologies such as Benz-diagrams and the V-topology. This has been done in 
\cite{Ablinger:2012sm}. 
%-------------------------------------------------------------------------------------------------
\begin{figure}[t]
\begin{center}
\includegraphics[scale=0.85]{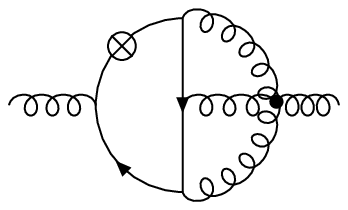}~~~
\includegraphics[scale=0.80]{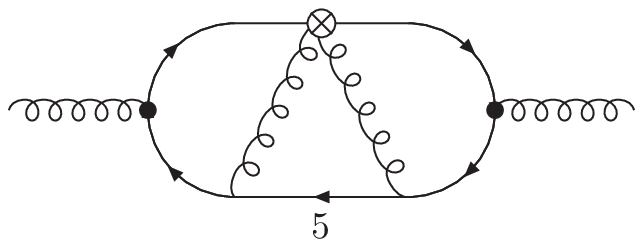}
\end{center}
\caption{An example of a diagram with Benz subtopology and a diagram of the 
V-topology}\label{FIG2}.
\end{figure}
%-------------------------------------------------------------------------------------------------
The diagram shown in Figure~\ref{FIG2} (left) results in 
%-------------------------------------------------------------------------------------------------
\begin{eqnarray}
I(N)&=&
\frac {1} {(N+1) (N+2)}
\Biggl\{
\frac{2 \left(1-13 (-1)^N+(-1)^N \textcolor{red}{2^{3+N}}+N-7 (-1)^N N+3 (-1)^N
    \textcolor{red}{2^{1+N}}N\right) }{(1+N) (2+N)} \zeta_3
\nonumber\\&&
+\frac{1}{(2+N)} S_3
+\frac{(-1)^N}{2 (2+N)} S_1^3
-\frac{(-1)^N (3+2 N)}{2 (1+N)^2 (2+N)} S_2
+\frac{5 (-1)^N}{2} S_2^2
\nonumber\\&&
+\frac{(-1)^N (3+2 N)}{2 (1+N)^2(2+N)} S_1^2
-\frac{(-1)^N}{2} S_2 S_1^2
+\frac{3 (-1)^N (4+3 N)}{(1+N) (2+N)} S_3
+3 (-1)^N S_4
+\frac{2}{(2+N)} S_{-2,1}
\nonumber\\&&
+{2 (-1)^N } \zeta_3 S_1 \left(2\right)
+\frac{2 (-1)^N (3+N) }{(1+N) (2+N)} S_{2,1}
-{12 (-1)^N } S_1 \zeta_3
\nonumber\\&&
+\frac{(-1)^N (5+7 N) }{2 (1+N) (2+N)} S_1 S_2
+{3 (-1)^N } S_1 S_3
+{4 (-1)^N } S_{2,1} S_1
-{4 (-1)^N} S_{3,1}
\nonumber\\&&
-\frac{4 \left((-1)^N \textcolor{red}{2^{2+N}}-3 \textcolor{red}{(-2)^N} N+3 (-1)^N 
\textcolor{red}{2^{1+N}} 
N\right)}{(1+N)
  (2+N)} \textcolor{green}{S_{1,2}\left(\frac{1}{2},1\right)}
-{5 (-1)^N } S_{2,1,1}
\nonumber\\&&
+\frac{2 \left(-(-1)^N \textcolor{red}{2^{2+N}}-13 \textcolor{red}{(-2)^N} N+5 (-1)^N 
\textcolor{red}{2^{1+N}}
    N\right)}{(1+N) (2+N)} \textcolor{green}{S_{1,1,1}\left(\frac{1}{2},1,1\right)}
\nonumber\\&&
-{2 (-1)^N } \textcolor{green}{S_{1,1,2}\left(2,\frac{1}{2},1\right)}
-{(-1)^N}  \textcolor{green}{S_{1,1,1,1}\left(2,\frac{1}{2},1,1\right)}
\Biggr\}~.
\nonumber
\end{eqnarray}
%-------------------------------------------------------------------------------------------------
Also in this case the asymptotic expansion is regular. The corresponding representation in 
$x$-space leads to generalized harmonic polylogarithms. In the case of the massive V-topology, 
cf. Figure~\ref{FIG2} (right), further extensions arise. Here finite nested binomial and inverse 
binomial sums weighted with generalized harmonic sums contribute. In $x$-space root-valued 
letters contribute to the alphabet, extending those of the harmonic polylogarithms by 30 letters 
in the case of the given graph. An example of a contributing sum is
%-------------------------------------------------------------------------------------------------
\begin{eqnarray}
\sum_{i=1}^N \frac{1}{(i+1) \displaystyle \binom{2i}{i}}\sum_{j=1}^i \binom{2j}{j} \frac{1}{j} 
S_2(j) &=& \int_0^1 dx \frac{x^N-1}{x-1} \Biggl[\frac{x}{2} \left(H^*_{\sf w_8,w_8,1,0}(x)
- \zeta_2 H^*_{\sf w_8,w_8}(x)\right) 
\nonumber\\ &&
- \frac{x}{\sqrt{x-1/4}}\left(H^*_{\sf w_8,1,0}(x)
- \zeta_2 H^*_{\sf w_8}(x)\right)\Biggr] 
\nonumber\\ &&
+\frac{2}{3} \zeta_3 \int_0^1 dx 
\frac{\left(\tfrac{x}{4}\right)^N-1}{x-4} \left[\frac{x}{2} H^*_{\sf w_3}(x) - 
\frac{x}{\sqrt{1-x}}\right],
\nonumber
\end{eqnarray}
%-------------------------------------------------------------------------------------------------
with the letters ${\sf w_3, w_8}$ given by
%-------------------------------------------------------------------------------------------------
\begin{eqnarray}
{\sf w_3} = \frac{1}{x \sqrt{1-x}},~~~~~~{\sf w_8} = \frac{1}{x \sqrt{x - 1/4}}.
\nonumber
\end{eqnarray}
%-------------------------------------------------------------------------------------------------
Here the harmonic polylogarithms $H^*$ are defined as iterated integrals w.r.t. the point 
$x=1$. In the case of the scalar integral of diagram Figure~\ref{FIG2} (right) potential 
divergencies 
$\propto 8^N, 4^N$
cancel, while the one $\propto 2^N$ remains. It is expected to cancel for the physical graphs.
%-------------------------------------------------------------------------------------------------
\section{$\mathbf{O(\alpha_s^2)}$ Charged Current Corrections}
%-------------------------------------------------------------------------------------------------

\noindent
Charged current data on heavy flavor production will improve the sea-quark densities. Therefore,
here the $O(\alpha_s^2)$ QCD corrections are desirable. In the present analyses \cite{Alekhin:2012ig} 
the $O(\alpha_s)$ contributions, cf.~\cite{Gluck:1996ve,Blumlein:2011zu}, are used. Since the 
charged current HERA data are located in the high $Q^2$ region, the asymptotic form of the 
$O(\alpha_s^2)$ corrections yields a sufficient representation. It has been studied in Ref. 
\cite{Buza:1997mg} before. Recently these corrections have been derived independently in 
\cite{BHP13} giving the representations both in Mellin and $x$-space, extending the former analysis 
and correcting some errors.
%%-----------------------------------------------------
\section{Calculation of OMEs containing Benz graphs}
%%-----------------------------------------------------

\vspace{1mm}\noindent
Recently we have calculated the massive 3-loop OMEs $A_{qq,Q}^{(3),\rm NS}$ and 
$A_{qq,Q}^{(3),\rm NS,TR}$ for general values of $N$ and obtained the Wilson coefficient 
$L_{qq,Q}^{(3),\rm NS}$, cf.~\cite{Bierenbaum:2009zt,Blumlein:2009rg}. The corresponding class 
of graphs contains also massive Benz diagrams. An extension of the code {\tt Reduze~2} 
\cite{Studerus:2009ye,vonManteuffel:2012np} to graphs with local operator insertions allowed to 
reduce the corresponding integrals to master integrals, which have been calculated using 
hypergeometric, Mellin-Barnes and advanced summation techniques \cite{SIGMA}. In course of this 
we have also computed the complete 2-loop anomalous dimensions for transversity 
$\gamma^{\pm,(1)}_{\rm NS,TR}$ \cite{TRANS2} and the contributions $\propto T_F$ of the 3-loop 
anomalous dimensions $\gamma^{\pm,(2)}_{\rm NS}$  and  $\gamma^{\pm,(2)}_{\rm NS,TR}$ in an ab 
initio calculation. In the first case we confirm the results of 
\cite{Larin:1993vu,Larin:1996wd,Retey:2000nq,Blumlein:2004xt,Moch:2004pa} and in the second
case our earlier moments \cite{Blumlein:2009rg} and the results in \cite{GRACEY,Bagaev:2012bw}. 
Details of this calculation are given in \cite{NS1}. The calculation of further massive OMEs is 
underway.

%%-----------------------------------------------------
\section{Conclusions}
%%-----------------------------------------------------

\noindent
Recently progress has been made towards the complete calculation of the 3-loop heavy
flavor corrections to DIS in the region $Q^2 \gg m^2$, including the matrix elements 
needed in the variable flavor number scheme at general values of $N$. The $O(n_f T_F^2 C_{F,A})$ 
contributions have been completed. The gluonic $O(T_F^2)$ terms are currently calculated, 
after all principal topologies have been solved. The renormalization in the 2-mass case has been 
performed and for all OMEs the moments $N = 2,4,6$ were calculated. Also the setup for a VFNS in 
case {\it both} charm and bottom become massless, has been derived. 
No hierarchy exists for these terms individually.
This scheme is different from the former single mass VFNS. Diagrams of ladder-, V- and 
Benz-topologies containing no singularities in $\varepsilon$ can be systematically calculated. 
Here new functions occur, including a larger number of root-letters in iterated 
integrals. All logarithmic contributions to the asymptotic heavy flavor Wilson coefficients  
have been determined \cite{Bierenbaum:2010jp}.
After the two Wilson coefficients $L_{qq,Q}^{(3), \rm ps}$ and $L_{qg,Q}^{(3)}$ had 
been computed in \cite{Ablinger:2010ty} we have calculated  $L_{qq,Q}^{(3),\rm NS}$ and 
$A_{qq,Q}^{(3),\rm NS,TR}$ as well as the associated 2- and 3-loop anomalous dimensions.
The calculation of further Wilson coefficients is underway.

%------------------------------------------------------------------------

%-----------------------------------------------------------------------------

\begin{thebibliography}{99}
%------------------------------------------------------------------------
%
%[1]
\bibitem{Alekhin:2012vu}
  S.~Alekhin, J.~Bl\"umlein, K.~Daum, K.~Lipka and S.~Moch,
  %``Precise charm-quark mass from deep-inelastic scattering,''
  Phys.\ Lett.\ B {\bf 720} (2013) 172.
%  [arXiv:1212.2355 [hep-ph]].
  %%CITATION = ARXIV:1212.2355;%%
%------------------------------------------------------------------------
%
%[2]
\bibitem{Buza:1995ie}
  M.~Buza, Y.~Matiounine, J.~Smith, R.~Migneron and W.~L.~van Neerven,
  %``Heavy quark coefficient functions at asymptotic values $Q~2 \gg m~2$,''
  Nucl.\ Phys.\  B {\bf 472} (1996) 611.
%  [arXiv:hep-ph/9601302].
  %%CITATION = NUPHA,B472,611;%%
%-----------------------------------------------------------------------------
%
%[3]
\bibitem{Blumlein:2006mh}
  J.~Bl\"umlein, A.~De Freitas, W.~L.~van Neerven and S.~Klein,
  %``The Longitudinal Heavy Quark Structure Function F**Q anti-Q(L) in the Region Q**2 >> m**2 at O(alpha**3(s)),''
  Nucl.\ Phys.\ B {\bf 755} (2006) 272.
%  [hep-ph/0608024].
  %%CITATION = HEP-PH/0608024;%%
%-----------------------------------------------------------------------------
%
%[4]
\bibitem{Vermaseren:2005qc} 
  J.~A.~M.~Vermaseren, A.~Vogt and S.~Moch,
  %``The third-order QCD corrections to deep-inelastic scattering by photon   
  %exchange,''
  Nucl.\ Phys.\  B {\bf 724} (2005) 3
%  [arXiv:hep-ph/0504242]
  %%CITATION = NUPHA,B724,3;%%
and refences therein. 
%-----------------------------------------------------------------------------
%
%[5]
\bibitem{Buza:1996wv}   
  M.~Buza, Y.~Matiounine, J.~Smith and W.~L.~van Neerven,
  %``Charm electroproduction viewed in the variable-flavour number scheme
  %versus fixed-order perturbation theory,''
  Eur.\ Phys.\ J.\  C {\bf 1} (1998) 301.
%  [arXiv:hep-ph/9612398].
  %%CITATION = EPHJA,C1,301;%%
%-----------------------------------------------------------------------------
%
%[6]
\bibitem{Bierenbaum:2009zt}
  I.~Bierenbaum, J.~Bl\"umlein and S.~Klein,
  %``The Gluonic Operator Matrix Elements at O(alpha(s)**2) for DIS Heavy Flavor Production,''
  Phys.\ Lett.\ B {\bf 672} (2009) 401.
  [arXiv:0901.0669 [hep-ph]].
  %%CITATION = ARXIV:0901.0669;%%
%------------------------------------------------------------------------
%
%[7]
\bibitem{Bierenbaum:2009mv}
  I.~Bierenbaum, J.~Bl\"umlein and S.~Klein,
  %``Mellin Moments of the O(alpha**3(s)) Heavy Flavor Contributions to unpolarized Deep-Inelastic Scattering at Q**2 >> m**2 and Anomalous Dimensions,''
  Nucl.\ Phys.\ B {\bf 820} (2009) 417.
%  [arXiv:0904.3563 [hep-ph]].
  %%CITATION = ARXIV:0904.3563;%%
%------------------------------------------------------------------------
%
%[8]
\bibitem{Blumlein:2009rg}
  J.~Bl\"umlein, S.~Klein and B.~T\"odtli,
  %``O(alpha(s)**2) and O(alpha(s)**3) Heavy Flavor Contributions to Transversity at Q**2 >>m**2,''
  Phys.\ Rev.\ D {\bf 80} (2009) 094010.
%  [arXiv:0909.1547 [hep-ph]].
  %%CITATION = ARXIV:0909.1547;%%
%------------------------------------------------------------------------
%
%[9]
\bibitem{Steinhauser:2000ry}
  M.~Steinhauser,
  %``MATAD: A program package for the computation of massive tadpoles,''
  Comput.\ Phys.\ Commun.\  {\bf 134} (2001) 335.
%  [arXiv:hep-ph/0009029].
  %%CITATION = CPHCB,134,335;%%
%-----------------------------------------------------------------------------
%
%[10]
\bibitem{Bierenbaum:2010jp}
  I.~Bierenbaum, J.~Bl\"umlein and S.~Klein,
  %``Logarithmic $O(\alpha_s^3)$ contributions to the DIS Heavy Flavor Wilson Coefficients at $Q^2 \gg m^2$,''
  PoS DIS {\bf 2010} (2010) 148
  [arXiv:1008.0792 [hep-ph]];\\
  %%CITATION = ARXIV:1008.0792;%%
I.~Bierenbaum, J.~Bl\"umlein, S.~Klein, and F. Wi\ss{}brock, to appear.
%------------------------------------------------------------------------
%
%[11]
\bibitem{Bierenbaum:2007qe}
  I.~Bierenbaum, J.~Bl\"umlein and S.~Klein,
  %``Two-Loop Massive Operator Matrix Elements and Unpolarized Heavy Flavor Production at Asymptotic Values Q**2 >> m**2,''
  Nucl.\ Phys.\ B {\bf 780} (2007) 40.
  %[hep-ph/0703285 [HEP-PH]].
  %%CITATION = HEP-PH/0703285;%%
%------------------------------------------------------------------------
%
%[12]
\bibitem{Bierenbaum:2008yu}
  I.~Bierenbaum, J.~Bl\"umlein, S.~Klein and C.~Schneider,
  %``Two-Loop Massive Operator Matrix Elements for Unpolarized Heavy Flavor Production to O(epsilon),''
  Nucl.\ Phys.\ B {\bf 803} (2008) 1.
  %[arXiv:0803.0273 [hep-ph]].
  %%CITATION = ARXIV:0803.0273;%%
%------------------------------------------------------------------------
%
%[13]
\bibitem{Ablinger:2010ty}
  J.~Ablinger, J.~Bl\"umlein, S.~Klein, C.~Schneider and F.~Wi\ss{}brock,
  %``The O(\alpha_s^3) Massive Operator Matrix Elements of O(n_f) for the Structure Function F_2(x,Q^2) and Transversity,''
  Nucl.\ Phys.\ B {\bf 844} (2011) 26.
  %[arXiv:1008.3347 [hep-ph]].
  %%CITATION = ARXIV:1008.3347;%%
%------------------------------------------------------------------------
%
%[14]
\bibitem{Blumlein:2012vq}
  J.~Bl\"umlein, A.~Hasselhuhn, S.~Klein and C.~Schneider,
  %``The $O(\alpha_s^3 n_f T_F^2 C_{A,F})$} Contributions to the Gluonic Massive Operator Matrix Elements,''
  Nucl.\ Phys.\ B {\bf 866} (2013) 196.
  %[arXiv:1205.4184 [hep-ph]].
  %%CITATION = ARXIV:1205.4184;%%
%------------------------------------------------------------------------
%
%[15]
\bibitem{Ablinger:2012ej}
  J.~Ablinger et al., %, J.~Blumlein, A.~De Freitas, A.~Hasselhuhn, S.~Klein, C.~Raab, M.~Round and 
  %C.~Schneider {\it et al.},
  %``Three-Loop Contributions to the Gluonic Massive Operator Matrix Elements at General Values of N,''
  PoS LL {\bf 2012} (2012) 033
  [arXiv:1212.6823 [hep-ph]].
  %%CITATION = ARXIV:1212.6823;%%
%------------------------------------------------------------------------
%
%[16]
\bibitem{Ablinger:2011pb}
  J.~Ablinger et al., %, J.~Blumlein, S.~Klein, C.~Schneider and F.~Wissbrock,
  %``3-Loop Heavy Flavor Corrections to DIS with two Massive Fermion Lines,''
  arXiv:1106.5937 [hep-ph];
  %%CITATION = ARXIV:1106.5937;%%
%\bibitem{Ablinger:2012qj}
%  J.~Ablinger, J.~Blumlein, A.~Hasselhuhn, S.~Klein, C.~Schneider and F.~Wissbrock,
  %``New Heavy Flavor Contributions to the DIS Structure Function $F_2(x,Q^2)$ at $O(\alpha_s^3),''
  PoS RADCOR {\bf 2011} (2011) 031.
%  [arXiv:1202.2700 [hep-ph]].
%------------------------------------------------------------------------
%
%[17]
\bibitem{BW13}
J. Bl\"umlein and F. Wi\ss{}brock, in preparation.
%------------------------------------------------------------------------
%
%[18]
\bibitem{Ablinger:2012sm}
  J.~Ablinger et al. % J.~Bl\umlein, A.~De Freitas, A.~Hasselhuhn, S.~Klein, C.~Schneider and 
  %F.~Wissbrock,
  %``New Results on the 3-Loop Heavy Flavor Wilson Coefficients in Deep-Inelastic Scattering,''
  arXiv:1212.5950 [hep-ph];\\
  %%CITATION = ARXIV:1212.5950;%%
J. Ablinger, J. Bl\"umlein, C.Raab, C. Schneider and F. Wi\ss{}brock, DESY 13-063.
%------------------------------------------------------------------------
%
%[19]
\bibitem{Ablinger:2012qm}
  J.~Ablinger, J.~Bl\"umlein, A.~Hasselhuhn, S.~Klein, C.~Schneider and F.~Wi\ss{}brock,
  %``Massive 3-loop Ladder Diagrams for Quarkonic Local Operator Matrix Elements,''
  Nucl.\ Phys.\ B {\bf 864} (2012) 52.
  %[arXiv:1206.2252 [hep-ph]].
  %%CITATION = ARXIV:1206.2252;%%
%------------------------------------------------------------------------
%
%[20]
\bibitem{SIGMA}
C. Schneider, {J. Symbolic Comput.} {\bf 43} (2008) 611, \newblock 
[arXiv:0808.2543v1];
{Ann. Comb.} {\bf 9} (2005) 75; {J. Differ. Equations Appl. }{\bf 11} (2005) 799;
{Ann. Comb. } {\bf 14} (4) (2010), [arXiv:0808.2596]; Proceedings of the Workshop
{\sf Motives, Quantum Field Theory, and Pseudodifferential Operators}, held at the
Clay
Mathematics Institute, Boston University, June 2--13, 2008, Clay Mathematics
Proceedings {\bf 12} (2010) pp.~285, arXiv:0904.2323 [cs.SC]
%%-308,
Eds. A.~Carey, D.~Ellwood, S.~Paycha, S.~Rosenberg;
{S\'em.~Lothar. Combin.} {\bf 56} (2007) 1, Article B56b,  Habilitationsschrift JKU
Linz
(2007) and references therein;\\
%\cite{Ablinger:2010pb}
%\bibitem{Ablinger:2010pb}
  J.~Ablinger, J.~Bl\"umlein, S.~Klein, C.~Schneider,
  %``Modern Summation Methods and the Computation of 2- and 3-loop Feynman Diagrams,''
  {Nucl.\ Phys.\ (Proc.\ Suppl.)}\  {\bf 205-206 } (2010)  110. %-115.
  %[arXiv:1006.4797 [math-ph]].
%-----------------------------------------------------------------------------------
%
%[21]
\bibitem{CODE1}
C. Schneider, {\it Simplifying Multiple Sums in Difference Fields}, In: {\sf Computer Algebra in 
Quantum Field Theory: Integration, Summation and Special Functions}, J. Bl\"umlein, C. Schneider 
(ed.), Texts and Monographs in Symbolic Computation, (Springer, Wien, 2013), 
arXiv:1304.4134 [cs.SC];\\
%\bibitem{Blumlein:2012hg}
  J.~Bl\"umlein, A.~Hasselhuhn and C.~Schneider,
  %``Evaluation of Multi-Sums for Large Scale Problems,''
  PoS RADCOR {\bf 2011} (2011) 032.
%  [arXiv:1202.4303 [math-ph]].
  %%CITATION = ARXIV:1202.4303;%%
%------------------------------------------------------------------------
%
%[22]
\bibitem{HSUM}
%\bibitem{Vermaseren:1998uu}
  J.~A.~M.~Vermaseren,
  %``Harmonic sums, Mellin transforms and integrals,''
  Int.\ J.\ Mod.\ Phys.\  A {\bf 14} (1999) 2037;\\
%  [arXiv:hep-ph/9806280].
  %%CITATION = IMPAE,A14,2037;%%
%\bibitem{Blumlein:1998if}
  J.~Bl\"umlein and S.~Kurth,
  %``Harmonic sums and Mellin transforms up to two-loop order,''
  Phys.\ Rev.\  D {\bf 60} (1999) 014018.
%  [arXiv:hep-ph/9810241].
  %%CITATION = PHRVA,D60,014018;%%
%-----------------------------------------------------------------------------
%
%[23]
\bibitem{Remiddi:1999ew}
  E.~Remiddi and J.~A.~M.~Vermaseren,
  %``Harmonic polylogarithms,''
  Int.\ J.\ Mod.\ Phys.\ A {\bf 15} (2000) 725.
%  [hep-ph/9905237].
  %%CITATION = HEP-PH/9905237;%%
%-----------------------------------------------------------------------------
%
%[24]
\bibitem{Moch:2001zr}
  S.~Moch, P.~Uwer and S.~Weinzierl,
  %``Nested sums, expansion of transcendental functions and multiscale multiloop integrals,''
  J.\ Math.\ Phys.\  {\bf 43} (2002) 3363
  [hep-ph/0110083].
  %%CITATION = HEP-PH/0110083;%%
%-----------------------------------------------------------------------------
%
%[25]
\bibitem{Ablinger:2013cf}
  J.~Ablinger, J.~Bl\"umlein and C.~Schneider,
  %``Analytic and Algorithmic Aspects of Generalized Harmonic Sums and Polylogarithms,''
  arXiv:1302.0378 [math-ph], J. Math. Phys, in print.
  %%CITATION = ARXIV:1302.0378;%%
%-----------------------------------------------------------------------------
%
%[26]
\bibitem{Ablinger:2011te}
  J.~Ablinger, J.~Bl\"umlein and C.~Schneider,
  %``Harmonic Sums and Polylogarithms Generated by Cyclotomic Polynomials,''
  J.\ Math.\ Phys.\  {\bf 52} (2011) 102301.
%  [arXiv:1105.6063 [math-ph]].
  %%CITATION = ARXIV:1105.6063;%%
%-----------------------------------------------------------------------------
%
%[27]
\bibitem{Ablinger:2013hcp}
  J.~Ablinger,
  %``Computer Algebra Algorithms for Special Functions in Particle Physics,''
  arXiv:1305.0687 [math-ph];
  %%CITATION = ARXIV:1305.0687;%%
%\bibitem{Ablinger:2010kw}
%  J.~Ablinger,
  %``A Computer Algebra Toolbox for Harmonic Sums Related to Particle Physics,''
  arXiv:1011.1176 [math-ph].
  %%CITATION = ARXIV:1011.1176;%%
%-----------------------------------------------------------------------------
%
%[28]
\bibitem{Ablinger:2013jta}
  J.~Ablinger and J.~Bl\"umlein,
  %``Harmonic Sums, Polylogarithms, Special Numbers, and their Generalizations,''
  arXiv:1304.7071 [math-ph].
  %%CITATION = ARXIV:1304.7071;%%
%-----------------------------------------------------------------------------
%
%[29]
\bibitem{QEXP}
%\bibitem{Harlander:1997zb}
  R.~Harlander, T.~Seidensticker, M.~Steinhauser,
%  {\it Complete corrections of $O(\alpha \alpha_s)$ to the decay of the $Z$-boson into
%  bottom quarks},
   Phys.\ Lett.\  {\bf B426 } (1998)  125;\\ %-132.
%  [hep-ph/9712228];\\
%\bibitem{Seidensticker:1999bb}
  T.~Seidensticker,
%  {\it Automatic application of successive asymptotic expansions of Feynman diagrams},
  [hep-ph/9905298].
%-----------------------------------------------------------------------------------
%   
%[30]
\bibitem{Blumlein:1998sh}
  J.~Bl\"umlein and W.~L.~van Neerven,
  %``Heavy flavor contributions to the deep inelastic scattering sum rules,''
  Phys.\ Lett.\ B {\bf 450} (1999) 417.
%  [hep-ph/9811351].
  %%CITATION = HEP-PH/9811351;%%
%-----------------------------------------------------------------------------------
%   
%[31]
\bibitem{Brown:2008um}
  F.~Brown,
  %``The Massless higher-loop two-point function,''
  Commun.\ Math.\ Phys.\  {\bf 287} (2009) 925.
%  [arXiv:0804.1660 [math.AG]].
  %%CITATION = ARXIV:0804.1660;%%
%------------------------------------------------------------------------
%
%[32]
\bibitem{Alekhin:2012ig}
  S.~Alekhin, J.~Bl\"umlein and S.~Moch,
  %``Parton Distribution Functions and Benchmark Cross Sections at NNLO,''
  Phys.\ Rev.\ D {\bf 86} (2012) 054009.
%  [arXiv:1202.2281 [hep-ph]].
  %%CITATION = ARXIV:1202.2281;%%
%------------------------------------------------------------------------
%
%[33]
\bibitem{Gluck:1996ve}
  M.~Gl\"uck, S.~Kretzer and E.~Reya,
  %``The Strange sea density and charm production in deep inelastic charged current processes,''
  Phys.\ Lett.\ B {\bf 380} (1996) 171
   [Erratum-ibid.\ B {\bf 405} (1997) 391].
%  [hep-ph/9603304].
  %%CITATION = HEP-PH/9603304;%%
%------------------------------------------------------------------------
%
%[34]
\bibitem{Blumlein:2011zu}
  J.~Bl\"umlein, A.~Hasselhuhn, P.~Kovacikova and S.~Moch,
  %``$O(\alpha_s)$ Heavy Flavor Corrections to Charged Current Deep-Inelastic Scattering in Mellin Space,''
  Phys.\ Lett.\ B {\bf 700} (2011) 294.
%  [arXiv:1104.3449 [hep-ph]].
  %%CITATION = ARXIV:1104.3449;%%
%------------------------------------------------------------------------
%
%[35]
\bibitem{Buza:1997mg}
  M.~Buza and W.~L.~van Neerven,
  %``O (alpha-s**2) contributions to charm production in charged current deep inelastic lepton - hadron scattering,''
  Nucl.\ Phys.\ B {\bf 500} (1997) 301.
%  [hep-ph/9702242].
  %%CITATION = HEP-PH/9702242;%%
%------------------------------------------------------------------------
%
%[36]
\bibitem{BHP13}
J. Bl\"umlein, A. Hasselhuhn, and T. Pfoh, in preparation.
%------------------------------------------------------------------------
%
%[37]
\bibitem{Studerus:2009ye}
  C.~Studerus,
  %``Reduze-Feynman Integral Reduction in C++,''
  Comput.\ Phys.\ Commun.\  {\bf 181} (2010) 1293.
%  [arXiv:0912.2546 [physics.comp-ph]].
  %%CITATION = ARXIV:0912.2546;%%
%------------------------------------------------------------------------
%
%[38]
\bibitem{vonManteuffel:2012np}
  A.~von Manteuffel and C.~Studerus,
  %``Reduze 2 - Distributed Feynman Integral Reduction,''
  arXiv:1201.4330 [hep-ph].
  %%CITATION = ARXIV:1201.4330;%%
%------------------------------------------------------------------------
%
%[39]
\bibitem{TRANS2}
%\bibitem{Hayashigaki:1997dn}
  A.~Hayashigaki, Y.~Kanazawa and Y.~Koike,
  %``Next-to-leading order Q**2-evolution of the transversity distribution
  %h1(x,Q**2),''
  Phys.\ Rev.\  D {\bf 56} (1997) 7350;\\
%  [arXiv:hep-ph/9707208];\\
  %%CITATION = PHRVA,D56,7350;%%
%-----
%\bibitem{Kumano:1997qp}
  S.~Kumano and M.~Miyama,
  %``Two-loop anomalous dimensions for the structure function h1,''
  Phys.\ Rev.\  D {\bf 56} (1997) 2504;\\
%  [arXiv:hep-ph/9706420].
  %%CITATION = PHRVA,D56,2504;%%
%-----
%\bibitem{Vogelsang:1997ak}
  W.~Vogelsang,
  %``Next-to-leading order evolution of transversity distributions and  Soffer's
  %inequality,''
  Phys.\ Rev.\  D {\bf 57} (1998) 1886
%  [arXiv:hep-ph/9706511] 
and references therein.
  %%CITATION = PHRVA,D57,1886;%%
%------------------------------------------------------------------------
%
%[40]
\bibitem{Larin:1993vu}
  S.~A.~Larin, T.~van Ritbergen and J.~A.~M.~Vermaseren,
  %``The Next Next-To-Leading QCD Approximation For Nonsinglet Moments Of Deep  
  %Inelastic Structure Functions,''
  Nucl.\ Phys.\  B {\bf 427} (1994) 41.
  %%CITATION = NUPHA,B427,41;%%
%------------------------------------------------------------------------------$
%
%[41]
\bibitem{Larin:1996wd}
  S.~A.~Larin, P.~Nogueira, T.~van Ritbergen and J.~A.~M.~Vermaseren,
  %``The Three loop QCD calculation of the moments of deep inelastic structure functions,''
  Nucl.\ Phys.\ B {\bf 492} (1997) 338.
%  [hep-ph/9605317].
  %%CITATION = HEP-PH/9605317;%%
%------------------------------------------------------------------------------$
%
%[42]
\bibitem{Retey:2000nq}
  A.~Retey and J.~A.~M.~Vermaseren,
  %``Some higher moments of deep inelastic structure functions at
  %next-to-next-to leading order of perturbative QCD,''
  Nucl.\ Phys.\  B {\bf 604} (2001) 281.
%  [arXiv:hep-ph/0007294]l.
  %%CITATION = NUPHA,B604,281;%%
%------------------------------------------------------------------------------$
%
%[43]
\bibitem{Blumlein:2004xt}
  J.~Bl\"umlein and J.~A.~M.~Vermaseren,
  %``The 16th moment of the non-singlet structure functions F2(x,Q**2) and
  %F(L)(x,Q**2) to O(alpha(s)**3),''
  Phys.\ Lett.\  B {\bf 606} (2005) 130.
%  [arXiv:hep-ph/0411111].
  %%CITATION = PHLTA,B606,130;%%
%------------------------------------------------------------------------------$
%
%[44]
\bibitem{Moch:2004pa}
  S.~Moch, J.~A.~M.~Vermaseren and A.~Vogt,
  %``The Three loop splitting functions in QCD: The Nonsinglet case,''
  Nucl.\ Phys.\ B {\bf 688} (2004) 101.
%  [hep-ph/0403192].
  %%CITATION = HEP-PH/0403192;%%
%------------------------------------------------------------------------
%
%[45]
\bibitem{GRACEY}
%\bibitem{Gracey:2003mr}
  J.~A.~Gracey,
  %``Three loop anomalous dimension of the second moment of the transversity operator in the MS-bar and RI-prime schemes,''
  Nucl.\ Phys.\ B {\bf 667} (2003) 242;
%  [hep-ph/0306163];
  %%CITATION = HEP-PH/0306163;%%
%\bibitem{Gracey:2006zr}
%  J.~A.~Gracey,
  %``Three loop anomalous dimensions of higher moments of the non-singlet twist-2 Wilson and transversity operators in the anti-MS and RI-prime schemes,''
  JHEP {\bf 0610} (2006) 040;
%  [hep-ph/0609231];
  %%CITATION = HEP-PH/0609231;%%
%\bibitem{Gracey:2006ah}
%  J.~A.~Gracey,
  %``Three loop MS-bar transversity operator anomalous dimensions for fixed moment n <= 8,''
  Phys.\ Lett.\ B {\bf 643} (2006) 374;
%  [hep-ph/0611071];
  %%CITATION = HEP-PH/0611071;%%
%\bibitem{Gracey:2007if}
%  J.~A.~Gracey,
  %``Three loop DIS and transversity operator anomalous dimensions in the RI-prime scheme,''
  PoS ACAT {\bf } (2007) 079.
%  [arXiv:0706.2071 [hep-ph]].
%------------------------------------------------------------------------
%
%[46]
\bibitem{Bagaev:2012bw}
  A.~A.~Bagaev, A.~V.~Bednyakov, A.~F.~Pikelner and V.~N.~Velizhanin,
  %``The 16th moment of the three loop anomalous dimension of the non-singlet transversity operator in QCD,''
  Phys.\ Lett.\ B {\bf 714} (2012) 76.
%  [arXiv:1206.2890 [hep-ph]].
  %%CITATION = ARXIV:1206.2890;%%
%------------------------------------------------------------------------
%
%[47]
\bibitem{NS1}
J. Ablinger et al., in preparation.
%------------------------------------------------------------------------
%-----------------------------------------------------------------------------
\end{thebibliography}
\end{document}